\documentclass[a4paper,12pt]{article}
\usepackage{bm}
\usepackage{amsmath}
\usepackage{amsfonts}
\usepackage{amssymb}
\usepackage{graphicx}
\usepackage{graphics}
\usepackage[dvips]{color}
\usepackage{anysize}

\newcommand{\be}{\begin{equation}}
\newcommand{\ee}{\end{equation}}
\newcommand{\ba}{\begin{eqnarray}}
\newcommand{\ea}{\end{eqnarray}}

\newcommand{\nk}{{\bf      k}}
\newcommand{\np}{{\bf      p}}
\newcommand{\nq}{{\bf      q}}

\newcommand{\npsi}{{\bf \npsi}}

\newcommand{\de}{\text{d}}

\newcommand{\non}{\nonumber}

\newcommand{\bma}{\begin{pmatrix}}
\newcommand{\ema}{\end{pmatrix}}

\title{Neutrino-induced single-pion production:\\ Kinematics and Cross Section\footnote{Proceedings of the 20th International Workshop on Neutrinos (NuFACT18), 12-18 August 2018 Blacksburg, Virginia, USA.}}

\author{Ra\'ul Gonz\'alez-Jim\'enez$^1$}

\date{\small{$^1$Grupo de F\'isica Nuclear, Departamento de Estructura de la Materia, F\'isica T\'ermica y Electr\'onica, Facultad de Ciencias F\'isicas, Universidad Complutense de Madrid, CEI Moncloa, Madrid 28040, Spain}}

\begin{document}
\maketitle

\begin{abstract}
In the energy range of present and future accelerator-based neutrino-oscillation experiments, single-pion production (SPP) is one of the main contributions to the neutrino-nucleus scattering cross section. 
For these neutrino energies, ranging from several hundreds of MeV to a few GeV, the SPP on the nucleus is usually described by the reaction in which the incoming lepton couples to one bound nucleon in the nucleus, producing a pion and the knock-out nucleon, along with the residual system and the scattered lepton in the final state.
Here, the kinematics and cross section formula for this process are discussed, although the formalism can be applied to other $2\rightarrow4$ processes.
\end{abstract}

\section{Kinematics}

\begin{figure}[htbp]
   \centering  
  \begin{minipage}{0.5\textwidth}
    \includegraphics[width=\linewidth,keepaspectratio=true]{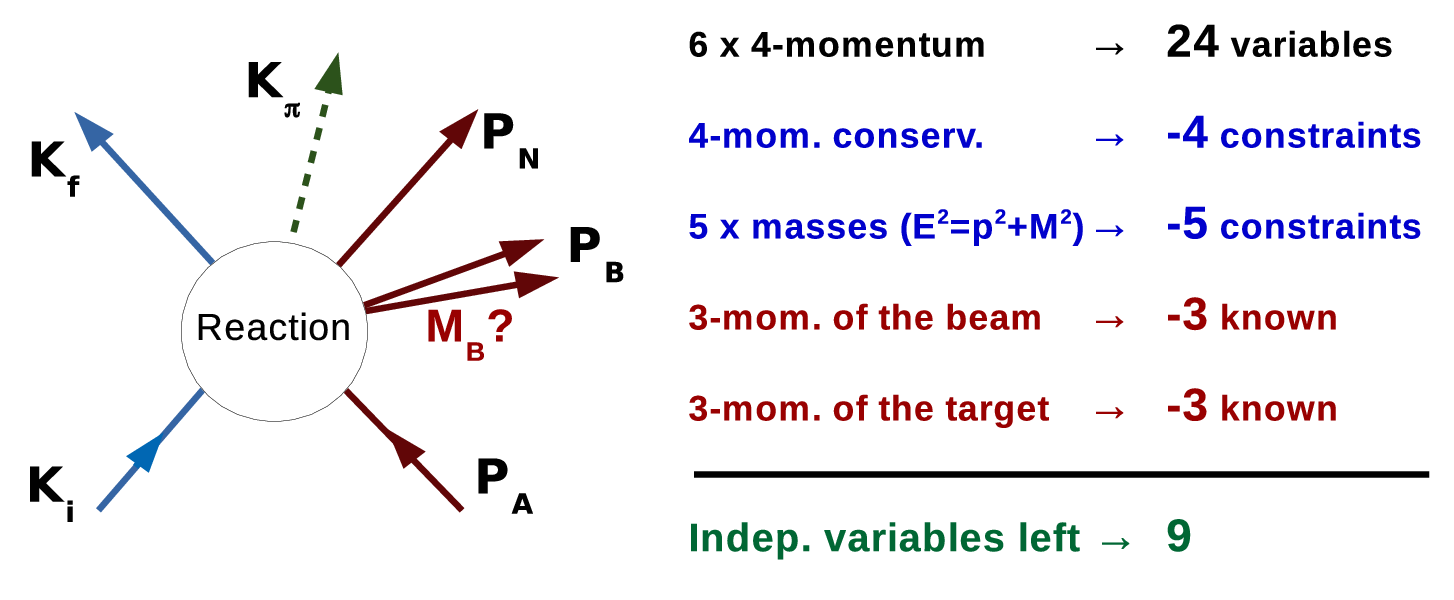}
  \end{minipage}
  \vline
  \begin{minipage}{0.5\textwidth}
    \scriptsize
    \begin{equation}\label{eA1}
      \begin{split}
	K_i^\mu &= (\sqrt{k_i^2+m_i^2},0,0,k_i)\,\\
	K_f^\mu &= (\sqrt{k_f^2+m_f^2},k_f\sin\theta_f,0,k_f\cos\theta_f)\,,\\  
	K_\pi^\mu &= (\sqrt{k_\pi^2+m_\pi^2},k_\pi\sin\theta_\pi\cos\phi_\pi,k_\pi\sin\theta_\pi\sin\phi_\pi,k_\pi\cos\theta_\pi)\,,\\
	P_N^\mu &= (\sqrt{p_N^2+M^2},p_N\sin\theta_N\cos\phi_N,p_N\sin\theta_N\sin\phi_N,p_N\cos\theta_N)\,,\\
	P_A^\mu &= (M_A,0,0,0)\,,\,\,P_B^\mu = (\sqrt{(E_m + M_A - M)^2 + p_B^2},\np_B).\non
      \end{split}
    \end{equation}
    \null
    \par\xdef\tpd{\the\prevdepth}
  \end{minipage}
  \vspace{-0.4cm}
  \caption{On the left hand side the SPP on the nucleus is sketched along with a summary of the i.v. needed to characterize the reaction. 
  On the right hand side, we list the 4-vectors in the lab frame written in terms of the i.v. The quantities $p_N$ and $\np_B$ are obtained by solving eq.~(\ref{pN1}). }
  \label{variables}
\end{figure}

First, let's figure out the number of independent variables (i.v.) needed to describe the scattering process shown in the left hand side of Fig.~\ref{variables}. This is a $2\rightarrow4$ process (2 incoming to 4 outgoing particles), which means 24 variables. Energy-momentum conservation
gives 4 constraints. 
The masses of all particles are known except that of the residual system (after the interaction the nucleus may be in an excited state or broken in pieces),
this provides 5 additional constraints through the energy-mass relation. 
%
Finally, we consider the 3-momentum of the incoming particles as fixed parameters: the target nucleus is at rest, the direction of the neutrino beam is known and its energy is averaged according to the flux distribution. Hence, the number of i.v. reduces to 9.  

We choose the lab variables $k_i$, $k_f$, $\theta_f$, $\phi_f$, $k_\pi$, $\theta_\pi$, $\phi_\pi$, $\theta_N$, $\phi_N$, and the missing energy $E_m$, as i.v. All the 4-vectors involved in the scattering can be written in terms of these ones (right hand side in Fig.~\ref{variables}). In particular, $p_N$ and $p_B$ are obtained as follows.
From energy and momentum conservation one has $\omega + M_A = E_B + E_N + E_\pi$ and $\np_B = \nq-\nk_\pi-\np_N$, with $\omega=\varepsilon_i-\varepsilon_f$ and $\nq=\nk_i-\nk_f$. Combining both equations:
\ba
 \omega + M_A = \sqrt{(E_m + M_A - M)^2 + (\nq-\nk_\pi)^2 + p_N^2 - 2\np_N\cdot(\nq-\nk_\pi)} + \sqrt{p_N^2+M^2} + E_\pi\,, \label{pN1}
\ea
where we have used $M_B = E_m + M_A - M$. Eq.~(\ref{pN1}) can be solved for $p_N$, and from that $p_B$ is trivially obtained (see Appendix~\ref{appendix}). It is interesting to point out that for some kinematics there are two physical solutions for $p_N$, each one will result in a different cross section. Thus, if e.g. the outgoing nucleon is not detected, these two cross sections should be added incoherently.

\section{Cross section}

We begin by describing the cross section formula for scattering off an off-shell nucleon. The residual nucleus will be introduced later in order to provide the correct energy-momentum balance. 
There are more rigorous (but complicated) ways of deriving the cross section formula that include corrections related to the center of mass~\cite{Udias1993}.

Starting from the general expression 
\ba
\de \sigma = \frac{|S_{fi}|^2}{T\phi_{inc}} \frac{V}{(2\pi)^3}\de \nk_f\ \frac{V}{(2\pi)^3}\de\nk_\pi\ \frac{V}{(2\pi)^3}\de\np_N
\ea
with $T$ a characteristic time of the system, $V$ the normalization volume,  $\phi_{inc}\approx1/V$ the incoming flux in the ultrarelativistic limit, and $S_{fi}$ the transition amplitude, it is straightforward to get 
%
\ba
  \frac{\de^{9}\sigma}{\de\nk_f\ \de\nk_\pi \de\np_N} = 
	\frac{1}{(2\pi)^{8}}\delta(E_N+E_\pi-\omega-E) {\cal F}\ l_{\mu\nu}h^{\mu\nu}\,,\label{cs1}
\ea
with $E$ the energy of the off-shell initial nucleon. The leptonic tensor $l_{\mu\nu}$ and the factor ${\cal F}$, that includes the boson propagator and coupling constants, were defined in Ref.~\cite{Gonzalez-Jimenez17}. The hadronic tensor $h^{\mu\nu}$ is model dependent and will not be discussed here (see, for example, Ref.~\cite{Gonzalez-Jimenez18}).\\

At this point, we account for the fact that the initial nucleon was bound in a nucleus, which also receives energy and momentum.  
The possibility that this residual nucleus is not in the ground state or, more generally, it is not a bound system, introduces one additional degree of freedom. This is incorporated in the formalism through the function $\rho(E_m)$ that represents the density of states of the residual nucleus. (As an example, in a pure shell model $\rho(E_m) = \sum_{shell}\delta(E_m-E_{shell})$ where $E_{shell}$ is the binding energy of each shell.)
Hence, Eq.~(\ref{cs1}) becomes
\ba
  \frac{\de^{10}\sigma}{\de\nk_f \de\nk_\pi \de\np_N\de E_m} = 
	\frac{1}{(2\pi)^{8}}\delta(E_N+E_\pi-\omega-M_A+E_B) \rho(E_m) {\cal F}\ l_{\mu\nu}h^{\mu\nu}\,,\label{cs2}
\ea
where the relation $E=M_A-E_B$ have been used in the Dirac delta. 
The energy-conservation delta can be used to integrate over $p_N$. We discuss here two possibilities:\\

i) {\it Neglecting the nuclear recoil}, i.e., $E_B\approx M_B=M_A-M+E_m$. One trivially obtains: 
\ba
\frac{\de^{9}\sigma}{\de E_f\de\Omega_f \de E_\pi\de\Omega_\pi \de\Omega_N\de E_m} = 
	\frac{\varepsilon_fk_f E_\pi k_\pi E_Np_N}{(2\pi)^{8}} \rho(E_m)\ {\cal F}\ l_{\mu\nu}h^{\mu\nu}\,.\label{CS_norec}
\ea
ii) {\it Considering the nuclear recoil}, so that $E_B = M_A-M+E_m + T_B$. In this case, the integration over $p_N$ gets more involved since $T_B$ depends on $p_N$. 
After some algebra one gets:
\ba
\frac{\de^{9}\sigma}{\de E_f\de\Omega_f \de E_\pi\de\Omega_\pi \de\Omega_N\de E_m} = 
	\frac{\varepsilon_fk_f E_\pi k_\pi E_Np_N}{(2\pi)^{8}\ f_{rec}} \rho(E_m)\ {\cal F}\ l_{\mu\nu}h^{\mu\nu}\,,
\ea
with $f_{rec} = \left|1 + \frac{E_N}{E_B}\left(1 +\frac{\np_N\cdot(\nk_\pi-\nq)}{p_N^2}\right) \right|$ the recoil factor.
Notice that in the limit $E_N/E_B<<1$ one gets $f_{rec}\rightarrow 1$ and the expression~(\ref{CS_norec}) is recovered.\\

To conclude, one could choose a different set of i.v. and then obtain a different cross section formula, but it should be clear that, for a given incoming energy, the cross section for the $2\rightarrow4$ process is necessarily a function of 8 i.v. (9 minus 1 azimuthal angle).
In current event generators, pion production usually starts off with the $2\rightarrow3$ process
$\nu_\ell + N \rightarrow \ell + N + \pi$, with $N$ a free nucleon,
and then build in nuclear effects. It would be important to make sure that the pion production model and the nuclear effects include awareness of the full spectrum of variables.\\

This work was supported by Comunidad de Madrid and Universidad Complutense de Madrid under the contract No. 2017-T2/TIC-5252.

\appendix

\section{Appendix}\label{appendix}

Here we provide the solution to Eq.~\ref{pN1} for $p_N$. 
We define the quantities:
\ba
  a_0   &\equiv& \omega + M_A - E_\pi\,,\\
  a_1   &\equiv& 2\hat\np_N\cdot(\nq-\nk_\pi)\non\\
          &=& (q^x-k_\pi^x)\sin\theta_N\cos\phi_N - k_\pi^y\sin\theta_N\sin\phi_N + (q^z-k_\pi^z)\cos\theta_N\,,\\
  a_2^2 &\equiv& (M_B^0+\epsilon_B^*)^2 + (\nq-\nk_\pi)^2\,.
\ea
In term of these, Eq.~\ref{pN1} reads:
\ba
  a_0 = \sqrt{a_2^2+p_N^2-2p_Na_1} + \sqrt{p_N^2+M^2}\,.
\ea
This is a second order equation in $p_N$, the two solutions are: 
\ba
  p_N = \frac{a_1a_3^2 \pm \sqrt{(a_1a_3^2)^2 - 4(a_0^2-a_1^2)(a_0^2M^2 - a_3^4/4)} }{2(a_0^2-a_1^2)}\,,
\ea
where $a_3^2\equiv a_0^2+M^2-a_2^2$. 
Notice that for some kinematics the two solutions are positive and therefore with physical meaning.

\end{document}